\documentclass{aastex701}
\usepackage{upgreek}

\begin{document}

\title{FRB Searches with the Irish LOFAR Station}

\author[0000-0001-7185-1310]{D. J. McKenna}
\affiliation{ASTRON, The Netherlands Institute for Radio Astronomy, Oude Hoogeveensedijk 4, 7991 PD Dwingeloo, The Netherlands}
\email{mckenna@astron.nl}

\author[0000-0002-4553-655X]{E. F. Keane}
\affiliation{School of Physics, Trinity College Dublin, College Green, Dublin 2, D02 PN40, Ireland}
\email{evan.keane@tcd.ie}

\begin{abstract} 

Here we report null results in the search for radio emission below
$200$~MHz from six known fast radio burst sources. The observations
reported here were taken using the Irish LOFAR station's high-band
antennas over the course of 2020, 2021 and 2022; the cumulative
observing time was $218$~h.

\end{abstract}


\section{Introduction}

The Irish LOFAR station (hereafter I-LOFAR) occassionally performs
searches for fast radio bursts (FRBs). Here we report null results
from such searches for $6$ targets. In each case the data were taken
using the high-band antennas in an observational setup identical to
that described in \citet{david_rrats}. Single-pulse searches using
\textsc{heimdall}~\citep{heimdall} with a very fine dispersion measure
tolerance~\citep{km26} of $1.005$ employed. The data remain available
as Stokes $I$ \textsc{sigproc}~\citep{sigproc} filterbanks, coherently
dedispersed at the known dispersion measure of each source to remove
intra-channel smearing, should they be of use in archival studies in
the future.

\section{Targets}

\subsection{SGR~1935+2154 (FRB~20200428A)}

The Galactic magnetar SGR~1935+2154 produced a roughly MJy flux
density burst in late April 2020 which was detected by both the CHIME
and STARE2 telescopes~\citep{Andersen2020,Bochenek2020}, and with
further follow-up multiple bursts were detected at several sites,
frequencies, and epochs in the following days and months (see
\citet{Rehan2023}, and many Astronomer's Telegrams, from
\citet{Zhu2020} to \citet{Maan2022}). The source was observed at
I-LOFAR for $17.5$~h between July 2020 and November 2022. Single-pulse
searches did not result in any credible detections across the span of
observations. The sessions did not overlap with any pulse
times-of-arrival reported by other telescopes.

\subsection{FRB~20180916A (R3)}

FRB~20180916A (aka `R3'), is an FRB source in a nearby spiral
galaxy~\citep{Andersen2019,Marcote2020}, and was the third known
repeating FRB source. It was found to do so on a 25\% duty cycle
across a roughly $16$-d period~\citep{Amiri2020}, making targeted
observations significantly easier than other sources. The source was
detected at $328$~MHz shortly after its
announcement~\citep{Pilia2020}, the lowest frequency detection of an
FRB at that time; this led to our attempts to detect it with I-LOFAR.
Observations were mainly performed between March 2020 and September
2020, for a cumulative $60.1$~h, with an additional $10$-h observation
in December 2020. No significant significant single pulse candidates
were detected. Following these observations a larger scale effort,
involving $223$ hours of observations with the LOFAR core was
undertaken by \citet{gop23} resulting in $11$ detected pulses. The
brightest of these would had a S/N of $\sim5$ at I-LOFAR (given the
nominal factor of $6$ relative sensitivity).

\subsection{FRB~20190303A (R17)}

A single observation of FRB~20190303A (aka `R17'), an off-disk FRB
source localised to a pair of colliding spiral
galaxies~\citep{Michilli2023}, was performed in March 2021. No pulses
were detected during the $4$-h observation.

\subsection{FRB~20200120E}

FRB~20200120E, sometimes referred to as the `M81 repeater' after its
host galaxy~\citep{Bhardwaj2021}, was observed in coordination with
other telescopes across multiple observing sessions from February to
June 2022. In total, the source was observed for $55.8$~h.  During
this time, $6$~h was spent observing with the Lovell telescope at
Jodrell bank (as described in \citealt{Rajwade2020}), and a further
$4.25$~h observation was performed with the HiPERCAM optical
instrument on the $10.4$-m Gran Telescopio
Canarias~\citep{hardy17,hipercam}. No observations resulted in
significant candidates from any of the telescopes.

\subsection{FRB~20201124A}

FRB~20201124A is a repeating FRB that was reported to be extremely
active both in the $400-800$~MHz CHIME
band~\citep{FRB20201124A_CHIME_ATel}, and at
$1.4$~GHz~\citep{Kirsten2021}; it was subsequently seen as low as
$325$~MHz~\citep{Main2022}. Assuming a flat or negative spectral
scaling (and ignoring scattering) it would be detectable with I-LOFAR
at the reported fluence levels. Observations between March 2021 and
March 2022, totalling $25.4$~h, did not result in any detectable
pulses.

\subsection{FRB~20220912A}

FRB~20220912A is a source of bright FRBs announced in October 2022
following the detection of 9 bursts in three
days~\citep{McKinven2022}. It was observed with I-LOFAR for $45.1$~h
between October 2022 and December 2022. This included a $12.8$-h
observation performed simultaneously with the Westerbork Synthesis
Radio Telescope, Stockert Radio Telescope and Torun Radio Telescope
reported in ~\citet{20220912A_ATel}. No pulses were detected during
any of these observations.

\section{Detection Limits}

Table~\ref{tab:limits} shows their flux density limits for each source.
These were calculated for a $10$-MHz band centred on $150$~MHz and
consider a $10\sigma$ detection threshold for a $10$-ms pulse; an
interested reader can scale these numbers in the usual way. The value
in the $145-155$~MHz range is quoted simply as this is a canonical
reference frequency. The sensitivity limit is within a few percent of
those quoted for each of the $125-135$, $135-145$, $145-155$ and
$155-165$~MHz ranges; it is somewhat worse as one goes towards the
band edges. 

\begin{table}[b!]
  \caption{Measured flux density limits for the 6 target sources observed in this work.}\label{tab:limits}
  \begin{center}
  \begin{tabular}{l|c}
    Source & $S_{150}$ limit (Jy) \\
    \hline
    SGR~1935$+$2154 (FRB~20200428A) & 25.9  \\
    FRB~180916 (R3) & 35.4 \\
    FRB~20190303A (R17) & 20.3 \\
    FRB~20200120E & 32.7 \\
    FRB~20201124A & 21.6 \\
    FRB~20220912A & 25.0 \\
    \hline    
  \end{tabular}
  \end{center}
\end{table}

\begin{acknowledgements}

I-LOFAR is situated at the Rosse Observatory, operated by Trinity
College Dublin. I-LOFAR infrastructure has benefited from funding from
Science Foundation Ireland, a predecessor of Taighde \'{E}ireann ---
Research Ireland.

\end{acknowledgements}

\begin{samepage}
\bibliographystyle{aasjournalv7}
\bibliography{rnaas_david_misc}
\end{samepage}

\end{document}